\documentclass{svjour3}
\smartqed  
\usepackage{graphicx}
\usepackage{amsmath}
\usepackage{amssymb}
\usepackage{subfigure}
\usepackage{color}
\usepackage{version}
\usepackage[latin1]{inputenc}
\titlerunning{Analytical solution of the $\mu(I)-$rheology\ldots}
\title{Analytical solution of the $\mu(I)-$rheology for fully developed granular flows in simple configurations}
\author{Merline Tankeo \and Patrick Richard and \'Edouard Canot}

\institute{Merline Tankeo 
\at Institut de Physique de Rennes, UMR CNRS 6251, Universit\'e de Rennes 1 - B\^atiment 11A, 35042 Rennes, France
\emph{Present address:} Universit\'e de Yaound\'e 1, Facult\'e des Sciences, D\'epartement d'informatique  B.P. 812 Yaound\'e, Cameroun
\and Patrick Richard 
\at LUNAM Universit\'e, IFSTTAR, Site de Nantes, Route de Bouaye, CS4, 44344 Bouguenais Cedex, France 
\at Institut de Physique de Rennes, UMR CNRS 6251, Universit\'e de Rennes 1 - B\^atiment 11A, 35042 Rennes, France
\and
\'Edouard Canot 
\at IRISA-CNRS, 
Universit\'e de Rennes 1
263 Avenue du Général Leclerc - B\^at 12
35042 RENNES Cedex, France
}

\date{}

\begin{document}\sloppy

\maketitle

\begin{abstract}

Using the $\mu(I)$ continuum model recently proposed for dense granular flows, we study  theoretically steady and fully  developed granular flows in two configurations: a plane shear cell and a channel made of two parallel  plates (Poiseuille configuration). In such a description, the granular medium behaves like a fluid whose viscosity is a function of the inertia. 
In the shear plane geometry our calculation predicts that the height of the shear bands scales with $U_0^{1/4}P_0^{1/2}$, where $U_0$ is the velocity of the moving plate and $P_0$ the pressure applied at its top.  In the Poiseuille configuration, the medium is sheared between the lateral boundaries and a plug flow is located in the center of the channel. The size of the plug flow is found to increase for a decreasing pressure gradient.
We show that, for small pressure gradient, the granular material behaves like a Bingham plastic fluid.
\end{abstract}

\section{Introduction}
Granular  flows~\cite{Delannay2007}  are of important scientific
interest because of their complex nature as well as  their wide occurrence
in industry and in environment. Unlike classical fluid
flows, they display different behaviors in different flow regimes
thus making difficult a complete and general constitutive
law from being derived.\\
Three flow regimes are generally reported in the literature.
In case of compact, slowly sheared
flows, the grains experience enduring contacts. They dissipate
energy by internal friction, so the constitutive law is
plastic-like. In case of dilute, rapidly sheared and agitated flows,
the granular materials interact mainly through collisions. The constitutive
law can be deduced from the kinetic theory of a gas of inelastic
grains~\cite{Jenkins1985}. In the intermediate flow regime, for example dense granular flows down an inclined plane, the granular
materials are dense as well as rapid, and subject to both
frictional and collisional stresses.\\
In the last decade, significant theoretical progresses~\cite{GDR2004,Cruz2005} has been made for   
 the latter regime. 
Those approaches consist in describing the granular medium as an incompressible fluid whose behavior is captured by a purely local rheology (called the $\mu(I)$ rheology) that can be used to write the stresses in  balance equations:

\begin{equation}
\left\{\begin{matrix}
\frac{\partial \mathbf{u}}{\partial t}+\left ( \mathbf{u}\cdot \boldsymbol\nabla \right )\mathbf{u}=-\frac{1}{\rho }\, \boldsymbol\nabla P +\frac{1}{\rho }\,\boldsymbol{\nabla} \cdot \boldsymbol{\tau},
\\ 
 \boldsymbol\nabla \cdot \mathbf{u}=0.
\end{matrix}\right.
\label{eqn:general}
\end{equation} 
In those equations, $P$ is the pressure, $\boldsymbol{\tau}$ the deviatoric  stress tensor, 
$\mathbf{u}$ the velocity and $\rho$ the bulk density. 
Such a rheology is able to reproduce observations from a great variety of experimental and numerical setups ~\cite{GDR2004,Cruz2005,Iordanoff2004,Jop2006a,Taberlet2007,Yohannes2010,Lagree2011}.
It is based on a coulombic friction model, and relates
the value of the effective coefficient of friction $\mu$ ({i.e.} the ratio of tangential to
normal stresses) to the non-dimensional inertial number $I$ that compares the typical time
scale of microscopic rearrangements with the typical time scale of macroscopic deformations:
%
\begin{equation}
  \frac{\left|\tau\right|}{P}=\mu \left ( I \right )\;\;\;\; \mbox{with}\;\;\;\; I=\frac{\left | \dot{\gamma } \right |d}{\sqrt{P/\rho_s}},
\label{eqn:muI}
\end{equation} 
where $\left|\tau\right|=\sqrt{\frac 12 \tau_{ij}\tau_{ij}}$ is the deviatoric stress tensor norm, $d$ is the particle diameter, $\rho_s$ is the particle density and $\dot\gamma$ the shear rate.
Note that the inertial number $I$ is the square root of the Savage number~\cite{Savage_AdApplMech_1984} also called the Coulomb number~\cite{Ancey_JoR_1999}.
%
%
It has been empirically shown \cite{Jop2005} that, for dense granular flows, the effective coefficient of friction $\mu$ of the system can be expressed by the following expression:
\begin{equation}
 \mu \left ( I \right )=\mu_s+\frac{\mu_2-\mu_s}{I_0/I+1}.
\label{eqn:muIcomplet}
\end{equation}

In the previous expression, $\mu_s$ is the threshold value for the quasi-static regime $(I \rightarrow 0)$. It corresponds to the angle of repose of the material. 
Therefore, the material flows only if the yield criterion $\left|\tau\right|>\mu_s P$ is satisfied. Below this threshold, the system behaves locally as a
rigid body. 
In strongly sheared regimes $(I \gg 1) $, $\mu (I)$ grows asymptotically towards $\mu_2$. 
In Eq.~(\ref{eqn:muIcomplet}), the values of the coefficients are material-dependent, for example the values for the spherical glass bead used in~\cite{Jop2005} are $\mu_s = \tan(20.9^\circ)$, $\mu_2=\tan(32.76^\circ)$ and $I_0=0.279$.
{If the inertial number $I$ is much lower than $I_0$ ($I\ll I_0$) the coefficient of friction can be approximated by the following  (simpler) expression: 
\begin{equation}
\mu(I)\approx \mu_s+(\mu_2-\mu_s)\frac{I}{I_0},\label{eqn:muIsimple}
\end{equation}
with $\mu_2>\mu_s$. 
{Recently it has been shown~\cite{Cortet_EPL_2009,Brodu_PRE_2013} that the tensorial extension~\cite{Jop2006a} of the $\mu(I)$ rheology is questionable since stress and strain tensors 
are not always aligned. Therefore the $\mu(I)$ should be applied only to monodirectional flows. In such a case Eq.~\ref{eqn:general} becomes
\begin{equation}
\frac{\partial u_x}{\partial t} + u_x \frac{\partial u_x}{\partial x} = -\frac{1}{\rho} \frac{\partial P}{\partial x} + \frac{1}{\rho}\frac{\partial \tau}{\partial y}, \\
\label{eqn:general_simple}
\end{equation}
}
where $\tau$ is the shear stress.\\
Another quantity of interest is the packing fraction $\Phi$, which has been found to decrease when the inertial number 
$I$ increases~\cite{Forterre2008}.
\begin{equation}
\Phi=\Phi_{max}-\zeta I,\label{eqn:compa}
\end{equation}
where $\Phi_{max}$ is the maximum packing fraction of the system and $\zeta$ is a positive constant typically equal to $0.2$. 
The latter equation is only valid for small values of $I$ since it leads to negative packing fractions for
$I>\Phi_{max}/\zeta$. This is consistent with the restriction of the $\mu(I)$ rheology to dense flows where relatively small values of $I$ are expected. However, for sake of simplicity,  
in the following, 
{ we will not take into account the latter equation and assume that the packing fraction does not depend on $I$: 
$\Phi=\Phi_{max}$.}
{This assumption will be discussed in the last section of the paper.}
For systems made of monodispersed spherical glass beads,  $\Phi_{max}\approx 0.6$.\\
As shown above, the $\mu(I)$-rheology is based on a phenomenological approach. Other models, based
on different theoretical backgrounds \cite{JenkinsBerzi2010,Aranson2001,Josserand2004,Mills1999,Louge2003,Berzi2011} can be found in the literature, but it has the advantage to be simple and to compare well against many experiments.
It should be however pointed out that this rheology is purely local, {i.e.}
the shear stress depends only on the local shear rate and pressure.
{Hence, it does not include long range correlations, which are prevalent near the jamming point~\cite{Ribiere2005b,Nichol2010,Reddy2011}.}
{A possible way to overcome this flaw, consists in introducing non-local effects in such models (see e.g.~\cite{Pouliquen_PTRSA_2009,Kamrin_PRL_2012}).} Another questionable point is that such a rheology does not use the notion of granular temperature which is at the base of the kinetic theory~\cite{Jenkins1985} even in the case of dense flows. Some discrepancies with experiments and simulations are also found in the case of dilute granular flows  or important inclination angles~\cite{Cortet_EPL_2009,Brodu_PRE_2013,Borzsonyi2009,Holyoake_JFM_2012}. 
{Moreover  the influence of the fluctuating energy flux is not taken into account. This point is problematic especially close to boundaries.} 
In spite of its flaws, the $\mu(I)$ rheology emerges so far as a reliable description of granular flows, at least if they are dense.\\
In this article, we use the $\mu(I)$ rheology to  solve analytically 
{the conservation of momentum} equation in the case of  two-dimensional granular flows in two simple setups: the shear plane and the Poiseuille configurations. We will also restrict ourselves to the cases of steady and fully developed flows, {\it i.e} flows whose properties depend neither on time nor on the position along the main flow axis.\\
The outline of this article is the following. In the next section we will present the assumptions used in this work. Section \ref{sec:resolution} is devoted to the presentation of the description of the analytical resolution that we used. Then, we will present the analytical approach, results and discussions for the shear plane flows (Sect.~\ref{sec:couette}) and the Poiseuille flows (Sect.~\ref{sec:poiseuille}). Finally we will present our conclusions.

\section{Simplifying assumptions: steady and fully developed flows}\label{sec:assumptions}
The analytical resolution of the Navier Stokes equations in the case of Newtonian fluids is difficult. The $\mu(I$) rheology introduces a non-constant viscosity that
complicate the resolution further. To bypass these difficulties, we restrict ourselves to the case of steady and fully developed granular flows.\\
Let us define the  $ x $-axis as the horizontal axis from left to right and the $ y $-axis as the vertical axis from bottom to top. The used assumptions are:
\begin{itemize}
  \item the flow is steady i.e. it no longer depends on time, which implies that ${\partial \mathbf{u}}/{\partial t} = 0 $.
  \item the flow is fully developed; that is, its properties (e.g. velocity) are nearly invariant along the main direction of flow. Consequently, we have for velocity, $ {\partial \vec{u}}/{\partial x}=0 $ (flow does not depend on $x$-direction), and then the $y$ component of the velocity is equal to zero, $u_y=0$.
  \item the pressure $P$ is supposed to be hydrostatic within the flow {i.e.} $ P = P_0 + \rho g (H-y) $ where $ H $ is the height of the flow and $P_0$ the external pressure. Note that this hypothesis was tested many times by simulation of discrete elements in different geometries \cite{Cruz2004,Silbert2001}.
\end{itemize}

Taking these assumptions into account, the system to solve (\ref{eqn:general_simple}) is reduced to the following differential equation:

\begin{equation}
 \frac{\partial }{\partial y}\left (\eta(y) \frac{\partial u }{\partial y} \right ) = K, 
\label{eqn:simplify}
\end{equation} 
where $ K ={\partial P }/{\partial x} $ is the pressure gradient in the direction of flow (assumed to be constant) and $\eta(y)= {\mu (I) P} / {\left | \dot {\gamma} \right |}$ is the effective dynamic viscosity. 
Let us recall here that the 
{ variations of the packing fraction are neglected }($\Phi=0.6$ uniformly within  the system).

\section{Analytical resolution}\label{sec:resolution}
We solve analytically 
the  nonlinear Eq. (\ref{eqn:simplify}). The steps of this calculation are the followings:

\begin{enumerate}
 \item We integrate analytically with respect to $y$ the pressure gradient ${\partial P}/{\partial x} $ (assumed to be constant). This leads to:
\begin{equation}
\displaystyle \eta(y) \frac{\partial u }{\partial y} =K\,y+k_1,
\label{eqn:1}
\end{equation} 
where $k_1$ is the constant of integration.

 \item We solve directly the Eq. (\ref {eqn:1}) with $ \dot\gamma = {\partial u} / {\partial y} $ as unknown. Let us recall here that $ \eta (y) =  {\mu (I) P} / {\left | \dot {\gamma} \right |} $ is the effective viscosity.

 \item We integrate with respect to $y$ the result $ {\partial u} / {\partial y} $ 
 and call $k_2$ the  constant of integration.

 \item By applying the boundary conditions of the studied  configuration, we obtain a system of two equations which allows the determination of the unknowns $ k_1 $ and $ k_2 $.
\end{enumerate}
As mentioned above, in the following two sections, we will apply this resolution to two simple configurations: the plane shear flow (Sect.~\ref{sec:couette}) and the Poiseuille flow (Sect.~\ref{sec:poiseuille}).

\section{Shear plane flow}\label{sec:couette}

We applied the 
{resolution described in Sect.~\ref{sec:resolution}} to the shear plane configuration (see Fig.~\ref{fig:couette}) with no pressure gradient ($K=0$). 
The granular medium is located between two plates separated by a height $ H $, $ P_0 $ is the pressure resulting from a vertical stress on the top plate, which moves at a constant velocity $ U_0 $, the lower one being fixed. This geometry has been studied intensively both experimentally~\cite{GDR2004} and numerically by discrete element methods~\cite{GDR2004,Cruz2005,Cruz2004}.

\begin{figure}[htb]
  \centering
  \includegraphics[scale=0.5]{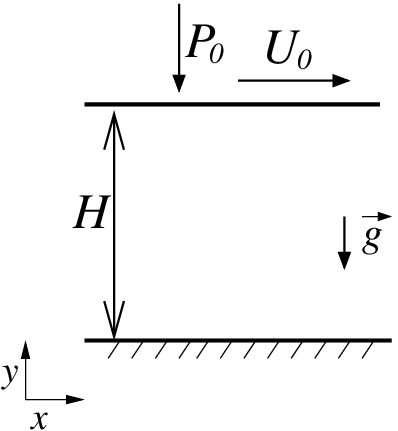}
  \caption{Sketch of the 2D shear plane configuration. $ H $ is the distance between the two plates, $ P_0 $ is the pressure resulting from a vertical stress on the top plate and $ U_0 $ the horizontal velocity of the same plate.}\label{fig:couette}
\end{figure}

\subsection{Dimensionless formulation}

In Eq. (\ref{eqn:simplify}), we have seven parameters that characterize the flow: $ U_0 $, $ P_0 $, $ H $, $ \rho $, $ \rho_s $, $g$ and $d$ which are respectively the velocity of the top plate, the pressure on the top plate, the height between the plates, the density of the granular medium, the grain density, the gravity and the diameter of the grains. To write Eq.~(\ref{eqn:simplify}) in  a dimensionless form, we must choose three scales: a length scale $ H $, a velocity scale $ U_0 $ and a pressure scale $ P_0 $.
The dimensionless variables are then:
$ \displaystyle y^*= {y}/{H} $, $ \displaystyle u^*= {u}/{U_0} $ and $ P^*= {P}/{P_0}$. 
From Vaschy-Buckingham theorem \cite{Buckingham1914} we can then reduce our set of parameters to only four dimensionless ones:

\begin{equation*}
\displaystyle \Lambda =\cfrac {H}{d} \, , \;\; \varepsilon =\cfrac{\rho g d}{P_0} \, , \;\; \alpha=\cfrac{U_0}{\sqrt{P_0/\rho_s}} \;\; \text{and} \;\; \Phi=\cfrac{\rho}{\rho_s}.
\end{equation*} 

The variables which depend on $ y ^ * $ are:
$ P^* (y^* )=1+\varepsilon \Lambda (1-y^* )$, $ I (y^* )=\cfrac{\alpha \left| {\dot \gamma}^* \right| }{\Lambda \sqrt{P^* (y^*)}}$, and $ \mu (I)= \mu_s + \cfrac{\mu_2-\mu_s}{I_0/I+1}$. Note that $\mu_s$, $\mu_2$ and $I_0$ are not considered as variables because they are constants of the $\mu (I)$-rheology. Thus, the dimensionless equation to solve is:
\begin{equation}
\displaystyle
\frac{\partial }{\partial y^*}\left [{\left| \dot{\gamma}^*\right|} \left( \varepsilon\,\Lambda\,\left( 1-y^*\right) +1\right) \,\left(\frac{ \alpha\left (\mu_2-\mu_s \right )}{\Lambda I_0\,\sqrt{\varepsilon\,\Lambda\,\left( 1-y^*\right) +1}\,+\alpha{\left| {\dot\gamma}^*\right|}}+\frac{\mu_s}{{\left| {\dot\gamma}^*\right|} }\right) \right ]=0.\label{eqn:dgammady}
\end{equation} 

In order to easily solve this equation, we must get rid of the absolute value that applies to the shear rate ${\dot\gamma}^*$. In our geometry, the top plate moves at a positive velocity and the bottom one is motionless. The shear rate within the granular system is therefore positive, or equal to zero:

\begin{equation}
{\dot{\gamma}}^* \geqslant 0 \;\;\; \text{then} \;\;\; {\left| \dot\gamma^*\right|}={ \dot\gamma^*}.
\label{eqn:gamma_positif}
\end{equation}

This allows us to obtain the following expression for the shear rate
\begin{equation}
\displaystyle
{ dot\gamma^*} = -\frac{\Lambda\,I_0\sqrt{\varepsilon \,\Lambda\left ( 1-y^{*} \right )+1}\left (\varepsilon \,\Lambda \,\mu_s \,y^*-\varepsilon \,\Lambda \,\mu_s -\mu_s+k_1 \right ) }{\alpha \left (\varepsilon \,\Lambda \,\mu_2 \,y^*-\varepsilon \,\Lambda \,\mu_2 -\mu_2+k_1 \right ). }
\label{eqn:gamma_couette}
\end{equation} 

where $k_1$ is the constant of integration of Eq.~(\ref{eqn:dgammady}).

The velocity is then obtained by integrating the shear rate with respect to $y^*$:

\begin{equation}
\displaystyle
\begin{matrix}
\displaystyle
u^*(y^*)= k_2 + \frac{{k_1}^{3/2}\,I_0\,\left ( \mu_s-\mu_2 \right )}{\alpha\,\varepsilon\,{ \mu_2}^\frac{5}{2}}\,\log\left (\frac{2\,\mu_2\,\sqrt{\Lambda\,\varepsilon\left ( 1-y^*\right )+1}- 2\,\sqrt{\mu_2\,k_1}}{2\,\mu_2\,\sqrt{\Lambda\,\varepsilon\left ( 1-y^*\right )+1}+ 2\,\sqrt{\mu_2\,k_1}} \right ) +
\\ 
\\
\displaystyle
\frac{2\,I_0}{3\,\alpha\,\varepsilon\,{\mu_2}^2} \left (\mu_2\,\mu_s\,\left(\varepsilon\,\Lambda\,\left ( 1-y^*\right )+ 1\right)^{\frac{3}{2}}\,+3\,k_1\left( \mu_s- \mu_2 \right) \sqrt{\varepsilon\,\Lambda\, \left ( 1-y^*\right )+ 1} \right )
\end{matrix}
\label{eqn:vitesse_couette}
\end{equation}

To completely define the velocity profile, it is necessary to determine the constants $k_1$ and $k_2$. This is done in the next section through the use of the boundary conditions.\\
%
%

\subsection{Boundary conditions}
Although the use of the $\mu(I)$ rheology close to boundaries, where the influence of the fluctuating energy flux may not be disregarded, is questionable we assume here that 
such an approximation does not modify significantly the features of the flow.  
{The relevancy of that assumption will be discussed in the last section of the present paper}.
Assuming  that  there is no slip on the walls, the velocity of the granular material at $y^*=1$ is equal to that of the top plate
  {i.e.} $ U_0 $. 
Since the pressure $P^*$ decreases with increasing $y^*$, we may observe situations where the yield criterion
$\tau/P > \mu_s$ is verified only if $y^*$ is larger than a critical value $y^*_{critical}$.
 Therefore, two situations have to be considered. First, the case where the yield criterion is verified at any depth.
In that case, 
  the velocity of the granular medium at the bottom plate is equal to the one of the bottom plate, {i.e.} zero. 
The corresponding boundary condition is therefore 
  $u^*(y^*=0)=0$.
The other situation is the case where the yield stress condition is only satisfied for $y^*\geq y^*_{critical}$. The flow is then localized close to the moving plate between
$y^*=1$ and $y^*_{critical}$.
In such a case, the former boundary condition is still valid but the latter   has to be replaced by $\tau(y^*=y^*_{critical}) / P^*(y^*=y^*_{critical}) = \mu_s$ and by $u^*(y^*=y^*_{critical})=0$.
From a practical point of view, $y^*_{critical}$, depends on $\alpha$ and $\varepsilon$, dependence which will be studied in the following. The flow localization is therefore observed only if the dimensionless height of the channel $\Lambda$ is greater than the dimensionless length $\Lambda_{critical}= \Lambda (1-y^*_{critical})$.
Note that, for given $\varepsilon$ and $\alpha$, if $\Lambda$ is set equal to $\Lambda_{critical}$, the conditions 
$\tau(y^*=y^*_{critical}) / P^*(y^*=\Lambda_{critical}) = \mu_s$ and $u^*(y^*=y^*_{critical})=0$ are equivalent to
$u^*(y^*=0)$.
\\
If the yield criterion is satisfied at any depth, the two boundary conditions allow us to find the values of $ k_1 $ and $ k_2 $ in step 4 of the resolution (see Sect.~\ref{sec:resolution}). To solve this nonlinear equation, we use the second boundary condition  and  Eq.~{(\ref{eqn:vitesse_couette})} to write  $ k_2 $ as a function of  $ k_1 $. Then, the same equation and the other boundary condition  are used to get the value of $ k_1$ by using Newton's iterative method. Note that it is necessary to choose an adequate initial value of $k_1$. Indeed, if we set $ f (k_1) = 0 $ the equation to be solved, the graph of $ f (k_1) $ has a vertical asymptote, which correspond to $I\rightarrow +\infty$, at the point of abscissa $ k_1 = \mu_2$, and no real values for $ k_1> \mu_2$. Practically, we therefore choose $ k_{1 \, \mbox {init}} = \mu_2 - 10^{-4} $.\\
If the flow is localized between $y^*=1$ and $y^*=y_{critical}^*$ the system composed of the three boundary conditions is solved numerically by using Newton's iterative method. 

\subsection{Results}
We have previously shown that the description of the flow depends on four parameters $\Lambda $, $\varepsilon $, $\alpha $ and $\Phi $. As mentioned above, 
{ the variations of the packing fraction $\Phi$ are neglected  within the flow}, so we restrict ourselves to the study of the influence of the other three parameters. Note that by definition, these parameters are all positive and different to zero except $ \varepsilon $ which can be zero if gravity is not taken into account. In that case, the shear rate $ \left |{\dot\gamma}^* \right |$ (see Eq.~(\ref{eqn:gamma_couette})) reduces to a constant $\left |{\dot\gamma}^* \right |= -\cfrac{\Lambda\,I_0\left ( k_1-\mu_s \right )}{\alpha \left ( k_1-\mu_2 \right )}$ and the velocity profile becomes linear with $y$, as follows:
$$u^*\left ( y^* \right ) =-\cfrac{\Lambda\,I_0\left ( k_1-\mu_s \right )y^*}{\alpha \left ( k_1-\mu_2 \right )}+k_2.$$

In the general case ($\varepsilon\neq 0$), we assigned values to the variables $ \varepsilon $ and $ \alpha $, and vary $ \Lambda $. Experimentally that 
corresponds to a variation of the height between the plates, or a variation of the grain diameter.
We chose values of $ \alpha $ and $ \varepsilon $ compatible with typical experimental situations on glass beads.
Thus, by choosing $\rho=1.5 \times 10^{3} \mbox{ kg/m}^3 $, $g=9.81 \mbox{ m/s}^2$, $d=0.5 \times 10^{-3} \mbox{ m}$, $P_0=1,000 \mbox{ Pa}$ and $U_0=100 \mbox{ mm/s}$ we obtain $\varepsilon=0.15 $ and $\alpha =0.007$.

Figure \ref{fig:vitesse_Couette} shows the velocity profiles for different values of $ \Lambda $. Note that for very small values of $ \Lambda $ ($ \Lambda \leq 2 $), the velocity profile tends to be linear, whereas for larger values ($ \Lambda = 5 \mbox { and } $ 10) the velocity profile is more curved.
If the gap between plates are even greater {i.e.} $\Lambda>\Lambda_{critical}$, the yield criterion is not satisfied between $y^*=0$ and 
$y^*=y^*_{critical}$ leading to a localization of the flow between the latter depth  and $y^*=1$.
Therefore, as expected, three cases can be observed:
\begin{itemize}
\item for $ \Lambda=\Lambda_{\rm{critical}} $ the granular system flows at any height and the shear rate is equal to zero at the bottom plate,
\item for  $ \Lambda <\Lambda_{\rm{critical}} $  the flow also occurs at any height but the shear rate is strictly positive at the the bottom plate,
\item for $ \Lambda >\Lambda_{\rm{critical}} $ the flow is localized. The static zone corresponds to $ y^* \in \left [0,1 -  {\Lambda_{\rm{critical}}} /{\Lambda} \right [$) and the flowing zone to 
$ y ^ * \in \left [1 -  {\Lambda_{\rm{critical}}} / {\Lambda}, 1 \right] $).
\end{itemize}
Note that, practically, the value of $ \Lambda_{\rm{critical}}$ is determined numerically by dichotomy on a given interval of $ \Lambda $.
Interestingly, as long as $\Lambda_{\rm{critical}}$ is defined ({i.e.} $\Lambda\geq\Lambda_{\rm{critical}}$),   it 
does not depend on $\Lambda$.
This result is in agreement with  experiments~\cite{GDR2004,Amarouchene2001,Komatsu2001}
as well as discrete element simulations~\cite{Cruz2005,Cruz2004} that report that under some conditions the flow in a plane shear cell is localized close to the moving surface. Below the aforementioned shear layers, the system is quasistatic. An increase of the height of the system modifies neither the  height of the shear layer nor the velocities of its grains. 
However, it should be noted that considering the zone below the shear layer as a purely static area is an approximation. Although it has been used many times~\cite{Amarouchene2001,Bouchaud1995,Boutreux1998,Taberlet2004b,Mangeney2007,Taberlet2008} it does not reflect the reality: the grains actually move intermittently ~\cite {Crassous2008,Richard2008} and the corresponding average profile decreases exponentially with depth. This discrepancy comes from the limitation of the $\mu (I)$-rheology that is not able to take into account the non-local effects responsible of the aforementioned intermittent motion.
\begin{figure}[htb]
  \centering
  \includegraphics[scale=0.6]{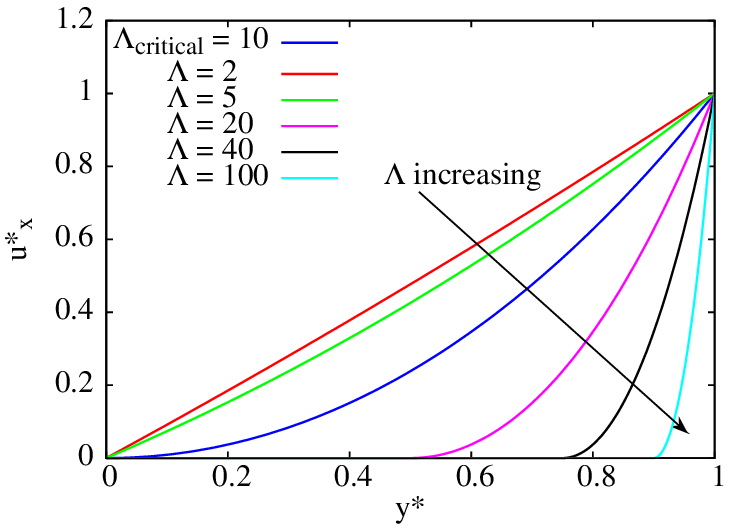}
  \caption{Velocity profiles obtained by varying $ \Lambda $ for $ \varepsilon = 0.15 $ 
  and $\alpha=0.007$. 
  For $ \Lambda = \Lambda_ {\rm{critical}} $, the shear rate is exactly zero at $y^*= 0 $. For $ \Lambda> \Lambda_{\rm{critical}} $ there is an area for which the velocity is equal to zero.}
  \label{fig:vitesse_Couette}
\end{figure}

\subsection{Discussion}

\subsubsection{Proposed law for $\Lambda_{\rm{critical}}$ }

We will now investigate the dependency of $ \Lambda_{\rm{critical}} $ with the other two parameters: $ \varepsilon $ and $ \alpha $. Figure \ref{fig:Lambda_alpha} reports the 
variations of $\Lambda_{\rm{critical}} $ versus $\alpha$ 
($\alpha \in [0.01 \,, \, 1]$) and for different values of $\varepsilon$.
\begin{figure}[htb]
\begin{center}
\includegraphics[scale=0.6]{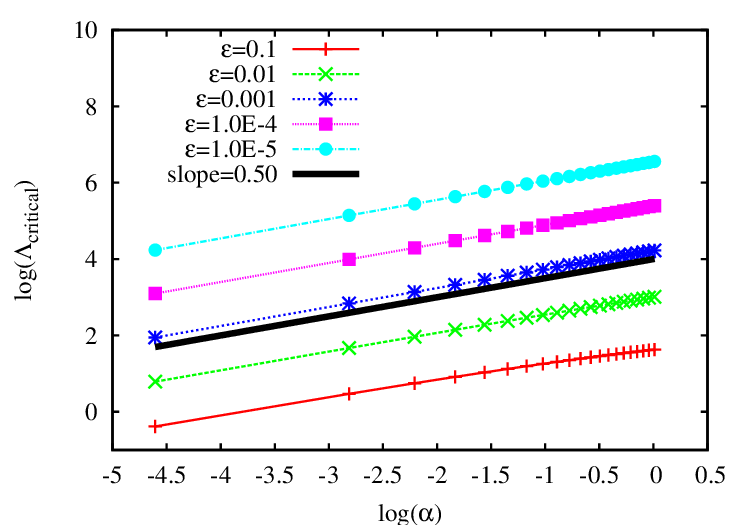}
\end{center}
\caption{Critical height $\Lambda_{\mbox{critical}}$ as a function of the dimensionless parameter $\alpha$ for several values of $\varepsilon$. For a given $\varepsilon$,  $\Lambda$ is proportional to $\alpha^{1/2}$. $\log$ refers to the neperian logarithm.}
\label{fig:Lambda_alpha}
\end{figure}
It shows that 
\begin{equation}
 \Lambda_{\rm{critical}} \approx f(\varepsilon)\,\alpha^{\frac{1}{2}},
\label{eqn:Lambda_alpha}
\end{equation} 
where $f(\varepsilon)$ is a function that describes the dependence of $\Lambda_{\rm{critical}}$ with
$\varepsilon$. 
Figure~\ref{fig:Lambda_epsilon} reports the variations of $ \Lambda $ for $ 1 / \varepsilon $ ranged from $ 0 $ to $ 100 $ and for different values of $ \alpha $. Let us recall that $1/\varepsilon$  is proportional to the pressure $P_0$. Therefore, studying the influence of $1/\varepsilon$ is equivalent to studying the effect of the external pressure $P_0$. We observe that 
\begin{equation}
 \Lambda_{\rm{critical}} \approx g(\alpha)\,1/\varepsilon^{\frac{1}{2}},
\label{eqn:Lambda_epsilon}
\end{equation} 
where $g(\alpha)$ is a function that describes the dependence of $\Lambda_{\rm{critical}}$ with
$\alpha$. 
\begin{figure}[htb]
\begin{center}
\includegraphics[scale=0.6]{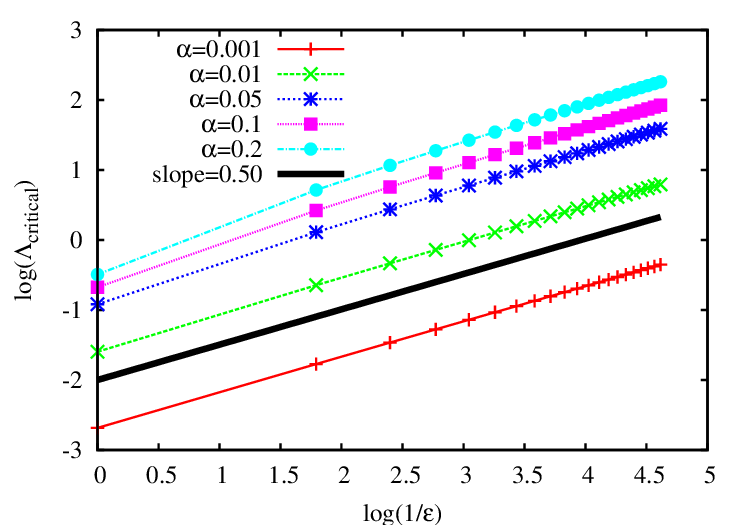}
\end{center}
\caption{Critical height $\Lambda_{\mbox{critical}}$ versus the dimensionless parameter $1/\varepsilon$ for several values of $\alpha$. For a given $\alpha$, 
$\Lambda$ is proportional to $1/\varepsilon^{1/2}$. $\log$ refers to the neperian logarithm.}
\label{fig:Lambda_epsilon}
\end{figure}
From equations (\ref{eqn:Lambda_alpha}) and (\ref{eqn:Lambda_epsilon}) we deduce:
\begin{equation}
 \Lambda_{\rm{critical}} \approx A \frac{\sqrt{\alpha}}{\sqrt{\varepsilon}},
\label{eqn:Lambda_epsilon_alpha}
\end{equation} 
where $A= 2.2 $ is calculated at point $\alpha=10^{-2}$, $\varepsilon=10^{-2}$.\\

An important point should be stressed out. In Figs.~\ref{fig:Lambda_alpha} and \ref{fig:Lambda_epsilon}, 
$\Lambda_{\rm{critical}}$ takes any value between $\approx 0$ and $\approx 10$.
However, since the height of a granular system cannot be smaller than the diameter of a grains, we have $H>d$ {i.e.} $\Lambda>1$. The latter condition and  relation (\ref{eqn:Lambda_epsilon_alpha})
lead to the following condition
$ 2.2 \sqrt {\alpha} \geqslant \sqrt {\varepsilon} $.
Figure~\ref{fig:Lambda_critique_carte} shows the interval of validity of the approached law given by  Eq.~(\ref{eqn:Lambda_epsilon_alpha}). We can observe that the aforementioned simplified equation does not hold for large values of $ \varepsilon $ and $ \alpha $ (for $ \log_ {10} (\varepsilon ^{-1 / 2}) <0.5 $ and $ \log_ {10} (\alpha ^{1 / 2})> $ -0.5). On the other side, the approximate law  better fits the exact results for small values of $ 1 / \varepsilon $ and $ \alpha $. 
That approached law as well as the relative deviations from that law --  whose isovalues are given by Fig.~\ref {fig:Lambda_critique_carte_err} --  will be discussed in next section. 

\begin{figure}[htbp]
 \centering
 \includegraphics[scale=0.45]{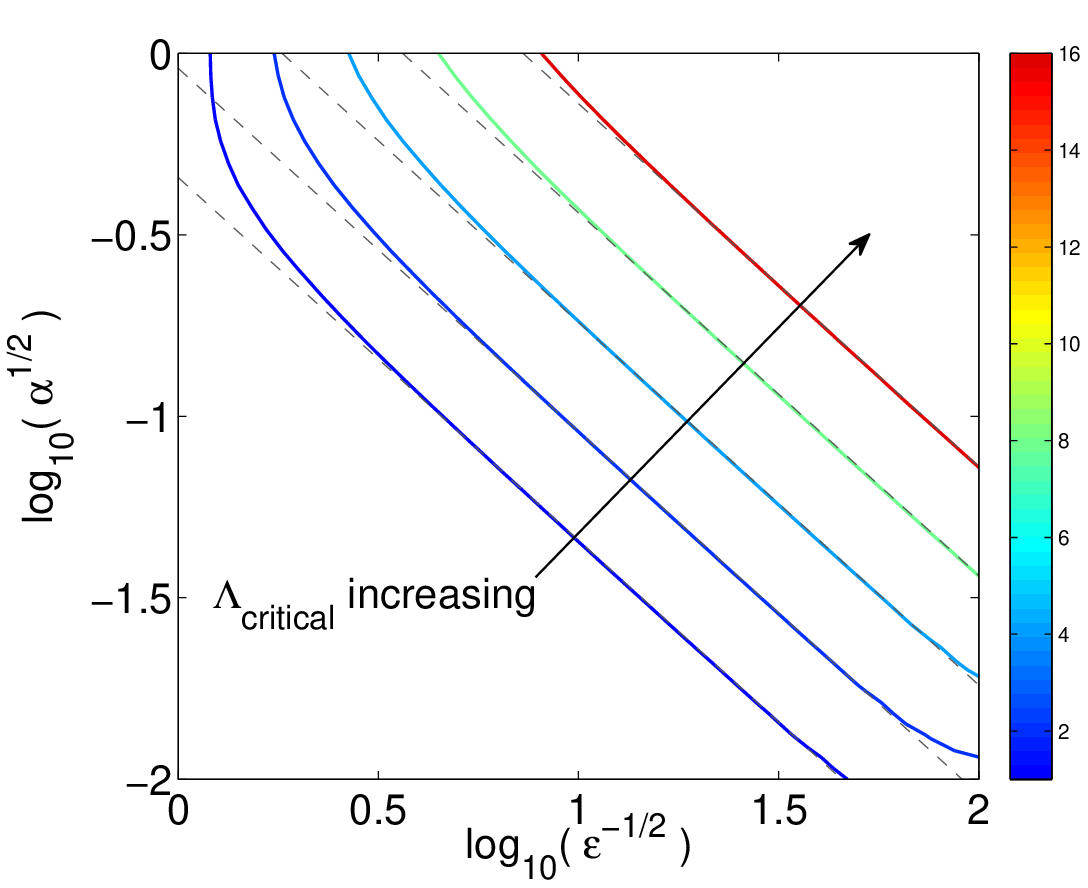}
 \caption{Isovalue lines of $ \Lambda_{\rm{critical}} = 1 ,\, 2,\, 4, \, 8, \, $ 16 (\textit{continuous line}), compared to the approximated law (\ref {eqn:Lambda_epsilon_alpha}) (\textit{dashed line}). The approximated law fits well the isovalue lines for values of $ \alpha $ and $ 1 / \varepsilon $ near to $\alpha=10^{-2}$, $\varepsilon=10^{-2}$. The map was obtained by a plot of isovalues of a matrix of 34 by 34 points.}
 \label{fig:Lambda_critique_carte}
\end{figure}

\begin{figure}[htbp]
 \centering
 \includegraphics[scale=0.45]{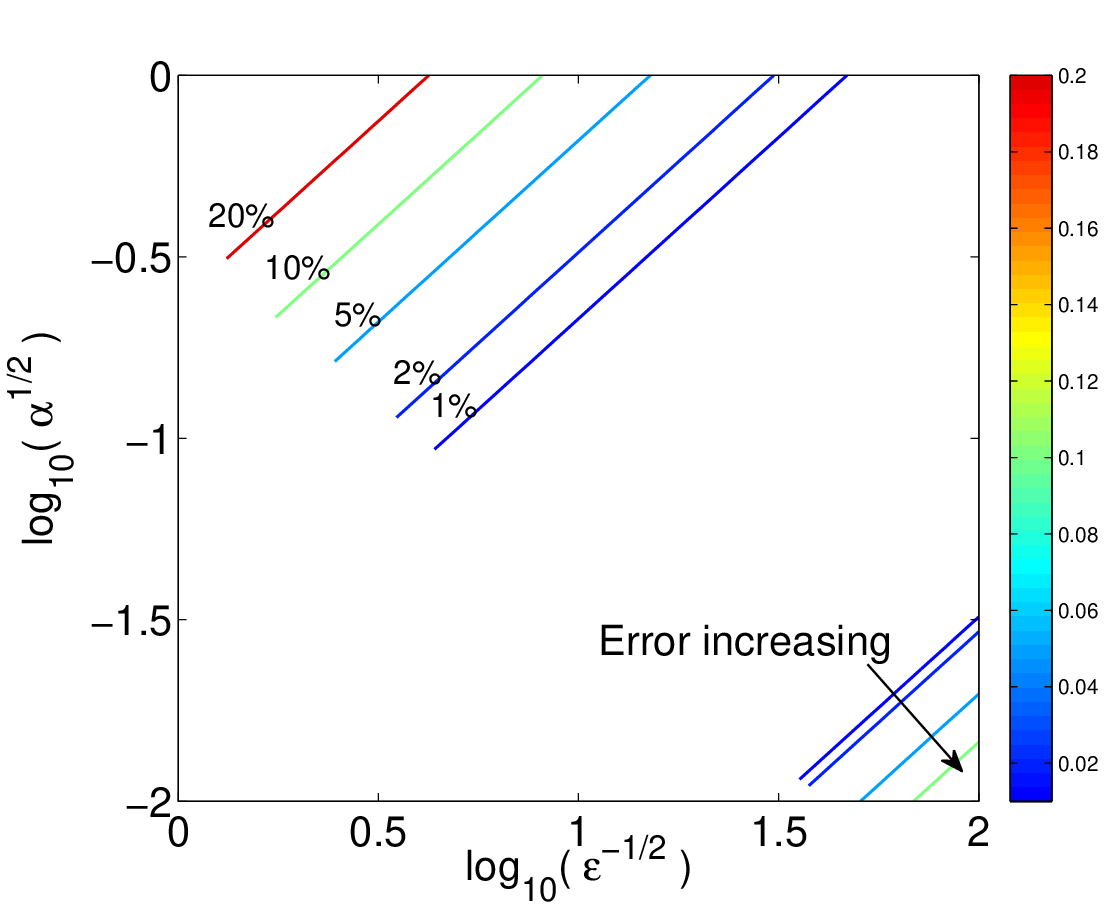}
 \caption{Isovalue lines of the relative error= 20\%, 10\%, 5\%, 2\%, 1\% of approximate law (\ref {eqn:Lambda_epsilon_alpha}). The map was obtained by a plot of isovalues of a matrix of 34 by 34 points.}
 \label{fig:Lambda_critique_carte_err}
\end{figure}

\subsubsection{Justification of the simplified law for $\Lambda_{\rm{critical}}$ }

In order to justify the dependency of $ \Lambda_{\rm{critical}} $ in $\sqrt {\alpha / \varepsilon} $, we will derive below this relationship by a simplified approach. That will allow us to better understand the origin of its domain of validity (see Fig.~\ref{fig:Lambda_critique_carte}).
In the configuration of the shear plane, the horizontal pressure gradient is zero $ \partial P (y) / \partial x = 0 $, the equation of motion therefore is $ \partial \tau (x, y) / \partial y =  0$, {i.e.} 
\begin{equation}
\frac{\partial \tau}{\partial y} = \frac{\partial \mu(y)P(y)}{\partial y} = 0.\label{eqn:mu_vs_P}
\end{equation}
After integration, we obtain
\begin{equation}
\mu(y) = \frac{A}{P(y)},
\end{equation}
 where $A$ is a constant.\\
One of the simplifying assumptions adopted (see Sect.~\ref{sec:assumptions}) is the hydrostatic character of the  pressure within the granular medium. Thus, in dimensionless form, we have
$P^*(y)=1 +\varepsilon \Lambda(1 - y^*)$. Then, we can write the Eq.~(\ref{eqn:mu_vs_P}) as
\begin{equation}
\mu(y^*)=\frac{A}{1+\varepsilon\Lambda(1-y^*)}.\label{eqn:mu_vs_P_bis}
\end{equation}
In the following, we will consider the case where the flow height is exactly equal to the critical height for which the shear stress becomes zero at the bottom plate (at $ y^*= 0 $). In this case, at this same plate, the granular system is at the limit of the static state and, consequently, we have $ \mu (y^*= 0) = \mu_s$.
Using this last relation in Eq.~(\ref {eqn:mu_vs_P_bis}) we can find the expression of the constant $ A $ and obtain for $ \mu (y ^ *) $ the following expression
\begin{equation}
\mu(y^*)=\mu_s\frac{1+\varepsilon\Lambda_{\rm{critical}}}{1+\varepsilon\Lambda_{\rm{critical}}(1-y^*)}.\label{eqn:mu_vs_P_fin}
\end{equation}
At $y^*=1$ this equation becomes
\begin{equation}
\mu(y^*=1)=\mu_s({1+\varepsilon\Lambda_{\rm{critical}}}).\label{eqn:mu_vs_P_fin_y1}
\end{equation}
We have an expression for the critical height $ \Lambda_{\rm{critical}} $ depending on the effective friction coefficient $ \mu (y^*)$. It is then sufficient to express the coefficient of friction as a function of $ \alpha $ and $ \varepsilon $ to derive an expression for $ \Lambda_{\rm{critical}} $ in terms of these quantities.
This can be done with the empirical formula connecting the effective friction with the inertial number $I $:
$\mu(I)=\mu_s+\left(\mu_2 - \mu_s \right)/({1+I_0/I }).$
The inertial number depends on $ y $ through pressure and shear rate:
$I(y^*)={\dot\gamma(y^*) \alpha }/({\Lambda_{\rm{critical}}\sqrt{P^*(y^*)}}).$

The pressure dependency with $y$  is known (hydrostatic assumption) contrary to that of the shear rate. To overcome this lack, we can assume that the velocity profile is linear between $ y^*= 1$ and $ y^*= 0 $.
In doing so, we underestimate the shear rate at the surface $ y^* =1$ but it seems reasonable to assume that this approximation does not alter the dependency of $ \dot \gamma $ in respect to $ \alpha $ and $ \varepsilon $.
Thus, within this approximation, the shear rate at the upper plate is $ \dot \gamma \approx U_0/H_{\rm{critical}}$ which corresponds to a dimensionless shear $ \dot \gamma^* $ of the order $ 1 $.
Moreover, still at $ y ^ *= 1 $, we have 
$P^*(y^*=1)=1$ and $I\approx {\alpha}/{\Lambda_{\rm{critical}}}$.
We can then deduce the following expression for $\mu(y^*)$:
$$\mu(y^*=1)\approx \mu_s+\left(\mu_2 - \mu_s \right)\frac{\alpha}{\alpha+I_0\Lambda_{\rm{critical}}}.$$
Then, this equation can be used to substitute $ \mu (y^*= 1) $ in the Eq.~(\ref{eqn:mu_vs_P_fin}) for which $y^*=1$ leading to the following second-order equation: 
\begin{equation}
\Lambda_{\rm{critical}}^2+\frac{\alpha}{I_0}\Lambda_{\rm{critical}}-\frac{\mu_2-\mu_s}{I_0\mu_s}\frac{\alpha}{\varepsilon} =0,\label{eqn:2ndordre}
\end{equation}
which has real solutions only if
\begin{equation}
\alpha\varepsilon \geq -4\left(\frac{\mu_2-\mu_s}{\mu_s}\right).
\end{equation}
Since the left hand side of this equation is negative, and the quantities $ \alpha $ and $ \varepsilon $ are positive, this condition is always satisfied. The only positive solution, physically acceptable, is therefore
\begin{equation}
\Lambda_{\rm{critical}} = 
\sqrt{\frac{\alpha}{\varepsilon}\frac{\mu_2-\mu_s}{\mu_sI_0}}\left(
\sqrt{1+\frac{\alpha\varepsilon\mu_s}{4I_0(\mu_2-\mu_s)}}-\sqrt{\frac{\alpha\varepsilon\mu_s}{4I_0(\mu_2-\mu_s)}}
\right)
\label{eqn:lambda_critique_2nd_ordre}
\end{equation} 
The dependency $\Lambda_{\rm{critical}}\propto \sqrt{\alpha/\varepsilon}$ is found if
$\zeta = \alpha\varepsilon\mu_s / [4I_0\left(\mu_2-\mu_s\right)] \ll 1.$ 
This condition corresponds to neglect the first order term in $ \Lambda_{\rm{critical}}$ in Eq.~(\ref{eqn:2ndordre}). In this case, we have
\begin{equation}
\Lambda_{\rm{critical}}=\sqrt{\frac{\mu_2-\mu_s}{\mu_sI_0}}\sqrt{\frac{\alpha}{\varepsilon}}\approx 1.57\sqrt{\frac{\alpha}{\varepsilon}}.
\label{eqn:Lambda_simplify}
\end{equation} 

Note the proximity of the coefficient $ 1.57 $ with the coefficient $ 2.2$ of Eq.~(\ref{eqn:Lambda_epsilon_alpha}). The weak difference comes from the approximation made above on the shear rate at the upper plate for $\Lambda=\Lambda_{\rm{critical}}$. By estimating the shear rate more accurately by using the Fig.~\ref{fig:vitesse_Couette} we find the right factor.\\

As mentioned above the law $\Lambda_{\rm{critical}}=2.2\sqrt{\alpha/\varepsilon}$ is an approximation and the isovalue lines of the relative error are reported in Fig.~\ref{fig:Lambda_critique_carte_err}.  The simple justification mentioned above is also able to explain the shape of those isovalues lines. Indeed, to obtain such a simple relation between $\Lambda_{\rm{critical}}$, $\alpha$ and $\varepsilon$ we have to assume that $\zeta$ is negligible with respect to $1$. 
If it is not the case, but if $\zeta$ is small, we can perform a first order Taylor expansion in $\zeta$ of Eq.~(\ref{eqn:lambda_critique_2nd_ordre}) that leads to
$$\Lambda_{\rm{critical}}\approx 
\sqrt{\frac{\alpha}{\varepsilon}\frac{\mu_2-\mu_s}{\mu_sI_0}}\left( 1-{\frac{\sqrt{\alpha\varepsilon}}{2\sqrt{I_0}}\sqrt\frac{\mu_s}{\mu_2-\mu_s}}\right).$$ 
Therefore, the relative error can be approximated by
$$\left|\Delta \Lambda_{\rm{critical}} / \Lambda_{\rm{critical}} \right| \approx
{\frac{\sqrt{\alpha\varepsilon}}{2\sqrt{I_0}}\sqrt\frac{\mu_s}{\mu_2-\mu_s}},$$
justifying why the isovalues  are more or less straight lines when they are plotted in the plane  ($\log\alpha^{1/2}$, 
$\log\varepsilon^{-1/2}$).

\section{Poiseuille flow}\label{sec:poiseuille}

The second configuration for which we applied the semi-analytical resolution is the Poiseuille flow. The granular medium flows in a channel, i.e. between two stationary plates (see Fig.~\ref{fig:poiseuille}) and a pressure difference between inlet and outlet of the channel is imposed. In this configuration we have been working in the absence of gravity. Here again we assume that the packing fraction is uniform within the flow and equal to $0.6$.
\begin{figure} [htb]
 \centering
 \includegraphics[scale=0.5]{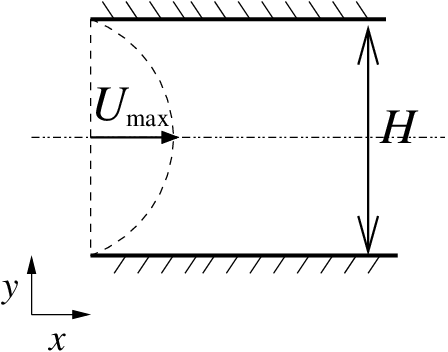}
 \caption{Sketch of the 2D Poiseuille configuration. $ H $ is the distance between the two plates and $ U_ {max} $ the maximum velocity of the imposed parabolic profile as input.}
 \label{fig:poiseuille}
\end{figure}

\subsection{Dimensionless formulation}
Following the same way as in the case of shear plane, to solve the Eq.~(\ref{eqn:simplify}), we have six parameters that characterize our flow, i.e. $ K $ [the pressure gradient, see Eq.~(\ref{eqn:1})], $ P_0 $, $ H $, $ \rho $, $ \rho_s $ and $ d $. Let us recall that gravity is not taken into account. To write Eq.~(\ref{eqn:simplify}) into a dimensionless form, we give three scales: a length scale $ H $, a velocity scale $ \sqrt {-K \, \frac{H}{\rho_s}} $ and a pressure scale $ P_0 $. The rescaled variables are then written as follows: $ \displaystyle y^*= \cfrac{y}{H} $, $ \displaystyle u^*= \cfrac {u}{\sqrt {-K \, \frac {H}{\rho_s }}}$, $ P^*= \cfrac {P}{P_0} $. This allows us to have three dimensionless variables:
\begin{equation*}
\displaystyle \Lambda =\cfrac {H}{d} \, , \;\; \beta=\cfrac{\sqrt{- K\,H } }{\sqrt{P_0}} \;\; \mbox{ and } \;\; \Phi=\cfrac{\rho}{\rho_s}. 
\end{equation*}

The variables which depends on $ y ^ * $ are:
$ P^* (y^* )=1$, $ I (y^* )=\cfrac{\beta \left|\dot \gamma^* \right| }{\Lambda \sqrt{P^* (y^* )}}$, and $ \mu (I)= \mu_s + \cfrac{\mu_2-\mu_s}{I_0/I+1}$. 

In the Poiseuille configuration we have a plane of symmetry (Fig.~\ref {fig:poiseuille}) which allows us to restrict our study to the half of the domain ($y^*\in[0,1/2]$) where the shear rate is positive $\dot{ \gamma}^* \geqslant 0$ then ${\left| \dot\gamma^*\right|}=\dot{ \gamma}^*$.

Thus the system is solved in dimensionless form:

\begin{equation}
\displaystyle
\frac{\partial }{\partial y^*} \left (\frac{{\beta\,{\left| \dot\gamma^*\right|}}\left (\mu_2-\mu_s \right )}{\Lambda\,I_0+\beta\,{\left| \dot\gamma^*\right|}}+\mu_s \right )=- \beta^2 ,
\end{equation} 

with

\begin{equation}
{\left| \dot\gamma^*\right|}= -\frac{\Lambda\,I_0\left ( \beta^2 \, y^*+\mu_s-k_1\right )}{\beta \left ( \beta^2 \, y^*+\mu_2-k_1\right )},
\label{eqn:gamma_poiseuille}
\end{equation} 
where $k_1$ is the constant of integration.
By integrating this equation with respect to $y^*$, 
{we  obtain, if the yield criterion is satisfied, the following expression of the velocity:
\begin{equation}
\displaystyle
u^*(y^*)=k_2-\frac{\Lambda\,I_0}{\beta^3}\,\left ( y^*\,\beta^2 + \left ( \mu_s-\mu_2 \right ) \log\left ( \beta^2\,y^*+\mu_2-k_1 \right ) \right ),
\label{eqn:vitesse_poiseuille}
\end{equation} 
where $ k_2 $ is the constant of integration.
\subsection{Boundary conditions}

 The two boundary conditions that we have in the half-Poiseuille configuration are:
\begin{itemize}
\item the shear stress is equal to zero at the center line, i. e. $\tau(y^*=1/2)=0$.
 \item we consider that the granular medium does not slide at the plate, i.e. $u^*(y^*)=0$ at $y^*=0$.
\end{itemize}

These two conditions allow us to find the value of $k_1$ and $k_2$ in step 4 of resolution (see Sect.~\ref{sec:resolution}), thus:
$$ k_1={\beta^2}/{2},$$
and
$$k_2=-\frac{\Lambda\,I_0\left ( \mu_2-\mu_s \right )}{\beta^3}\log\left ( \mu_2-\frac{\beta^2}{2} \right ). $$
The dimensionless shear stress is then given by $\tau^*=\beta^2(1/2 - y^*)$. 
Note that, since this quantity has an upper boundary $\mu_2$, a steady state cannot be attained, in the framework of this theory, if $\beta$ (the dimensionless pressure gradient) is higher than $\sqrt{2\mu_2}$. In the following we will therefore  consider that $\beta<\sqrt{2\mu_2}$.
{
Let us now determine $y_c^*$, the value of $y^*$ for which the yield criterion is no more satisfied
{i.e.} $\tau < \mu_sP$. From the latter expression of $\tau$ it corresponds 
to $y^*_c=1/2-\mu_s/\beta^2.$ Below this value, the yield criterion is satisfied and the system is sheared. Above, it behaves like  a plug flow.
Physically, $y^*_c$ cannot be lower than $0$. This conditions leads to  
$\beta^2>2\mu_s$. So, in the following, we will consider that $\beta \in [\sqrt{2\mu_s},\sqrt{2\mu_2}]$.}
After integration of the shear rate, 
{we  obtain for $y^*<y_c^*$,} the following expression of the velocity:

\begin{equation}
u^*(y^*)=-\cfrac{\Lambda\,I_0}{\beta^3}\,
\left (\beta^2\,y^*+
\left ( \mu_s-\mu_2 \right ) \log\left ( -\cfrac{2\,\beta^2\,y^*+2\, \mu_2-\beta^2}{\beta^2-2\, \mu_2} \right ) \right ).
\label{eqn:vitesse_poiseuille_complet}
\end{equation} 
The maximum value of the velocity is then 
\begin{equation}
u^*_{m}=u^*(y^*=1/2-\mu_s/\beta^2)=-\cfrac{\Lambda\,I_0}{\beta^3}\,
\left(\beta^2/2-\mu_s+
(\mu_s-\mu_2)\log(\frac{\mu_2-\mu_s}{\mu_2-\beta^2/2} )
\right).
\end{equation}
{Interestingly, the limit case $y^*_c=1/2$ is only obtained for infinite pressure gradient 
{i.e.} $\beta \to +\infty$ which is incompatible with the aforementioned condition 
$\beta\in [\sqrt{2\mu_s},\sqrt{2\mu_2}]$. 
Therefore, in such a geometry, the flow always displays a plug flow at the center of the cell. Its minimum and maximum sizes are respectively $\mu_s/\mu_2$ (obtained for $\beta\rightarrow \sqrt{2\mu_2}$) and 
$1$ (obtained for $\beta\rightarrow \sqrt{2\mu_s}$).}

\subsection{Results and Discussion}

In the current configuration, the description of the flow depends on two parameters $ \Lambda $ and $ \beta $ as 
{ the variations of $ \Phi $ are neglected } ($ \Phi = 0.6$). For a granular flow with the parameters $\rho = 1.5 \times 10^{3} \; \mbox{kg/m}^3 $, $ g = 9.81 \; \mbox {m/s}^2 $, $ d = 0.5 \times 10^{-3} \; \mbox{m} $, $ H = 0.1 \; \mbox {m} $, $ P_0 = 100 \; \mbox {Pa} $ and $ K =- 100 \; \mbox {Pa/m} $ we have $ \Lambda = 20 $ and $ \beta = 0.31$. We study the influence of these parameters on the velocity profile of the flow. Equation (\ref{eqn:vitesse_poiseuille_complet}) clearly shows that the amplitude of the velocity profile is proportional to $ \Lambda$. Thus, the study will be restricted to the influence of the parameter $ \beta $ ($\Lambda$ is kept constant and equal to $20$).\\
\begin{figure}[htb]
 \centering
 \includegraphics*[scale=0.30,angle=0]{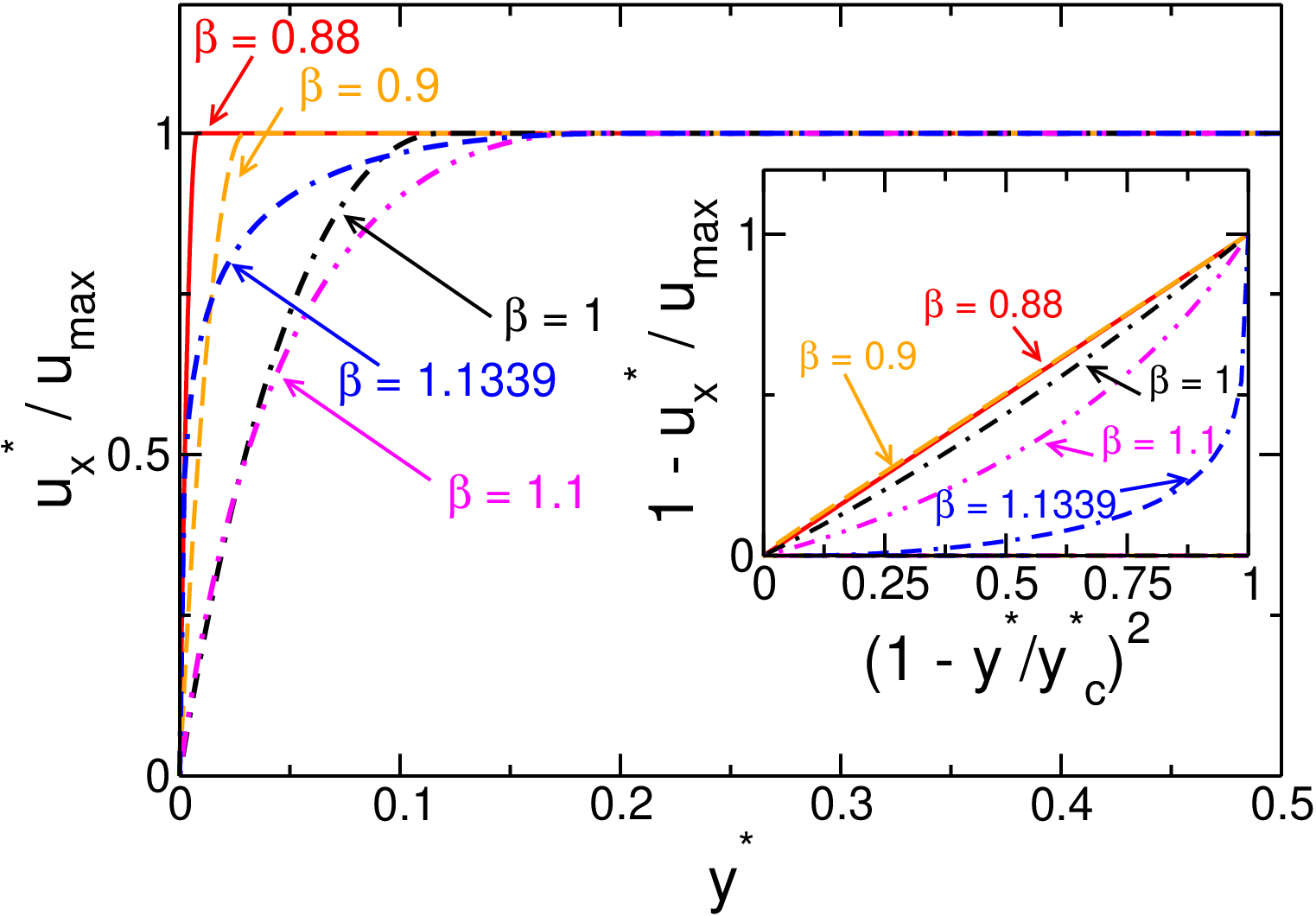}
 \caption{Variation of velocity profile for different $ \beta $ and for $ \Lambda = 20$. The Inset represents the profile of the sheared regions which tend to to be parabolic 
  for $ \beta \to \sqrt{2\mu_s}=0.8739$.}
 \label{fig:vitesse_poiseuille_beta}
\end{figure}

Figure~\ref{fig:vitesse_poiseuille_beta} reports the profiles obtained for different values of $ \beta $.
{As expected a plug flow is visible at the center of the channel for any values of 
$\beta\in [\sqrt{2\mu_s},\sqrt{2\mu_2}]$. 
Those profiles are somewhat close to the ones obtained by the 
``Da Vinci fluid'' model described in~\cite{Blumenfeld2011}.
They also look similar to the profiles obtained with Bingham plastic fluids, {i.e.} a
material that behaves as a rigid body at low stresses but flows as a viscous fluid at high stress.
Let us recall here that  Poiseuille flows of such fluids display a plug at the center of the cell and a parabolic velocity profile close to the sidewalls. 
To quantify this resemblance, we report in the inset of Fig.~\ref{fig:vitesse_poiseuille_beta},
the quantity $(1-u^*_x/u_{max})$ versus $(1-y^*/y^*_c)^2$. For Bingham plastic fluids, those two dimensionless quantities
are equal. 
Interestingly, we find that the velocity profiles correspond to those of a Bingham plastic fluid 
for small values of $\beta$ (e.g. $ \beta = 0.88, \mbox{and } 0.90)$. They flatten 
for larger values (e.g. $ \beta = 1, \; 1.1, \; 1.339$).}
{
Another quantity of interest is the maximum velocity $u^*_{m}$, which is equal to $ u ^ * (y^*=y^*_c) $.}
\begin{figure}[htb]
 \centering
 \includegraphics*[scale=0.30,angle=0]{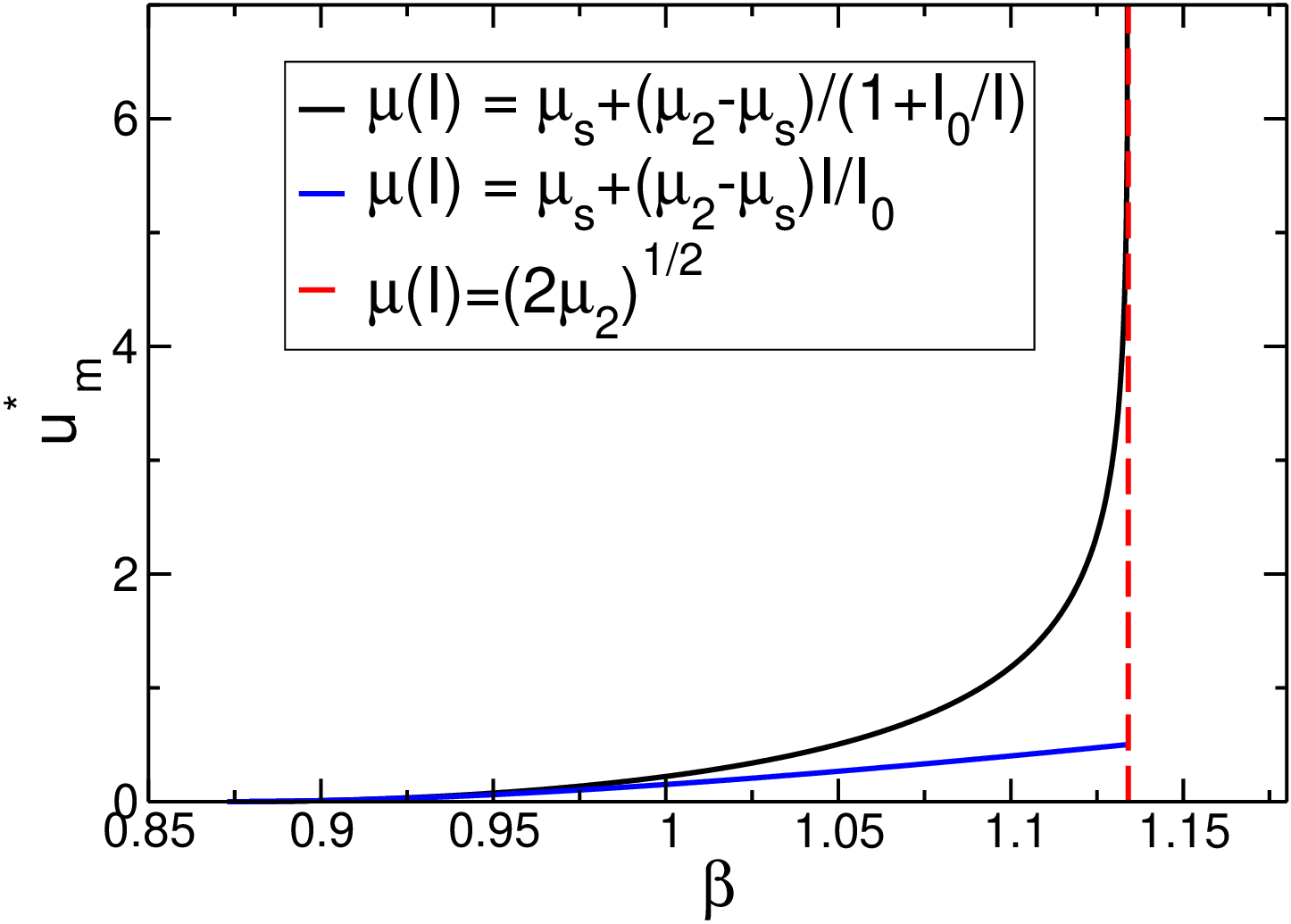}
\caption{Variation of $ u^*_{m} $ as a function of $ \beta $ and for $ \Lambda = $ 20. A Bingham plastic fluid and a granular material obeying the $\mu(I)$-rheology display a similar behavior  when $ \beta \to  0$. On the contrary, the latter displays  a divergence in velocity for when $ \beta \to \sqrt {2 \, \mu_2} = 1.1344 $  whereas the former remains linear.}
 \label{fig:U_max}
\end{figure}
{Figure~\ref{fig:U_max}, which reports previous quantity  
as a function of $\beta$, shows that $u_m^*$ is equal to zero for $\beta=\sqrt{2\mu_s}$ and diverges when $\beta$ tends to $\sqrt{2\mu_2}$.}
{The 
presence of this divergence can be explained as follows.
The balance of flow momentum over half the cell width gives that the pressure gradient $K$ times the half width $H/2$ is balanced by the difference in shear stresses on the wall and on the center of the cell. The former shear stress is $\left[\mu_s + (\mu_2-\mu_s)/(1+I_0/I)\right]P_0$ and the latter  is equal to zero. Consequently, $K H/2$ can increase no further than the limiting stress difference $\mu_2 P_0$ obtained as $I$ and $u^*_m$ become infinite. }



This figure also shows that for small value of $\beta $ ($\beta < 0.95 $), the velocity $U_{max} $ is the same as in the case of a 
Bingham plastic fluid ({i.e.} a parabolic profile between $0$ and $y_c^*$, then a plug flow) although the viscosity of the granular fluid is not that of a Bingham plastic fluid. Figure~\ref{fig:vitesse_poiseuille_beta} also shows that for the same range of $\beta$ ($\beta < 1 $) the velocity profile of the sheared region is approximated by a parabola.
To justify 
Bingham-like behavior for $\beta \to \sqrt{2\mu_s} $, let us first recall that for such fluid, when the yield criterion is satisfied, the shear stress is equal to 
to $\tau_c+\eta_n\dot\gamma$, where $\tau_c$ is the yield stress and $\eta_n$ a constant Newtonian viscosity. 
{Then, let us consider now the $\mu(I)$-rheology  and assume that $I$ is much smaller that $I_0$, justifying the approximation 
$\mu(I)=\mu_s+(\mu_2-\mu_s)I/I_0$. If the yield criterion is satisfied, the shear stress
is then Bingham-like:
%
\begin{equation}
\tau=\tau_c+\eta_n\left|\dot\gamma\right|,
\end{equation}
with $\tau_c=\mu_sP$ and $\eta_n=(\mu_2-\mu_s)\sqrt{\rho_SP}/I_0.$
In such a case, we obtain the following expressions of the velocity for $y^*\leq y^*_c:$
\begin{equation}
u^*=\frac{I_0\Lambda}{2\beta(\mu_s-\mu_2)}\left(\beta^2 {y^*}^2+(2\mu_s-\beta^2)y^*\right)
\end{equation}
Its maximum value is given by
\begin{equation}
u_m^*=u^*(y^*=y^*_c) = \frac{\Lambda I_0}{8\beta^3}\frac{(\beta^2-2\mu_s)^2}{\mu_s-\mu_2}.
\end{equation}
}

\begin{figure}[htb]
 \centering
 \includegraphics*[scale=0.30,angle=0]{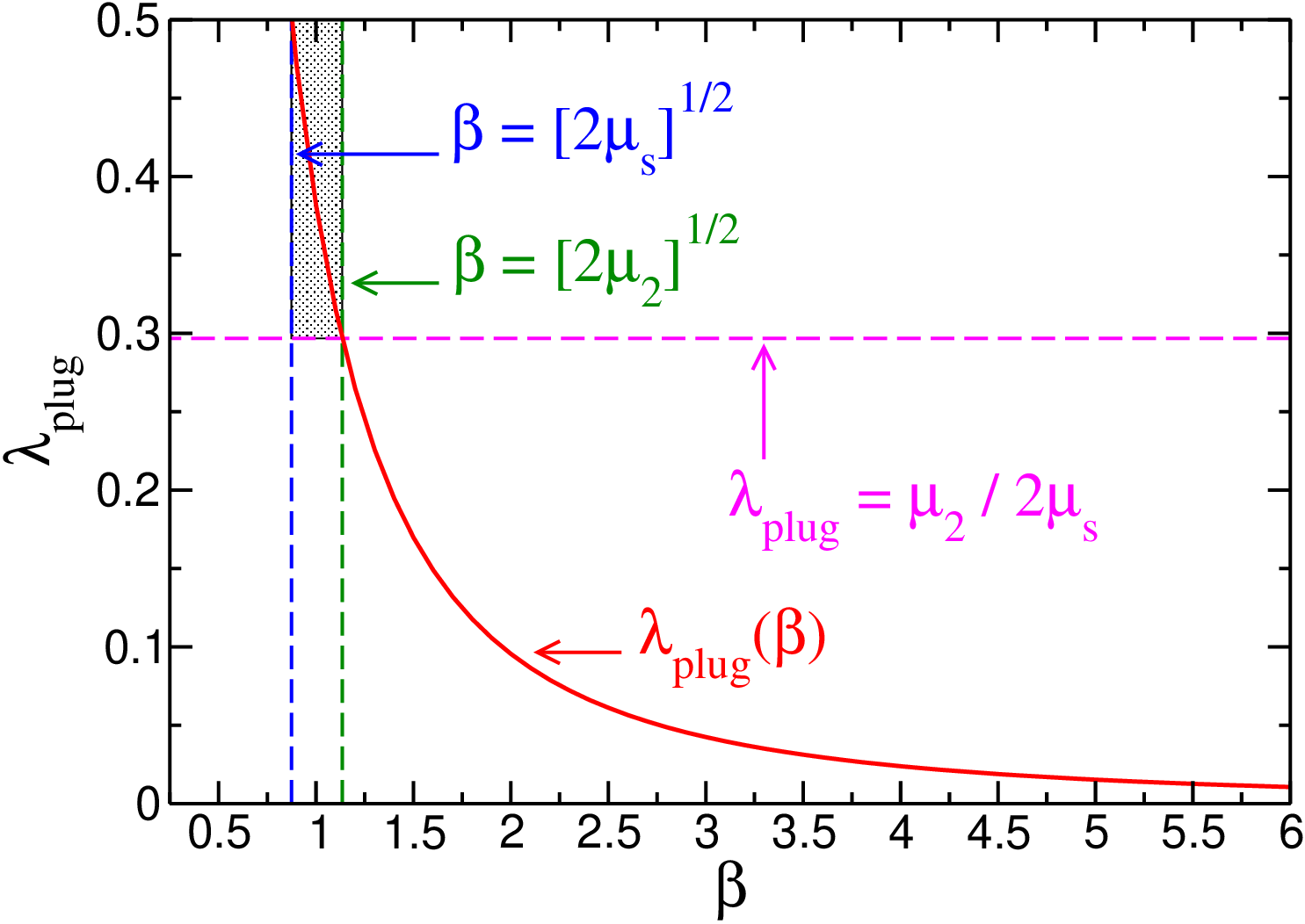}
\caption{The half-length of the plug flow in the Poiseuille configuration depends on the gradient pressure $\beta$. Depending on the expression used for $\mu(I)$ (see text for details), this half-length is between $0.5$ and $\mu_s/2\mu_2$ (full expression given by Eq.~\ref{eqn:muI} - gray zone) or $0.5$ and $0$ (simplified expression given by Eq.~\ref{eqn:muIsimple}).}
 \label{fig:lambda_plug}
\end{figure}
{It is worth noting that, if the expression used for the effective friction coefficient $\mu(I)$ is the simplified one (Eq.~\ref{eqn:muIsimple}), $\beta$ has no upper limit.  Therefore, the position of the plug flow $y^*_c$ belongs to the range $[0,1/2]$. Then, when $\beta\rightarrow\infty$  the half-length of the plug flow,
$\lambda_{\rm{plug}}$, tends to zero and the system is sheared all along its length. 
This point is illustrated on Fig.~\ref{fig:lambda_plug} where $\lambda_{\rm{plug}}$ is reported versus 
$\beta$. The gray zone corresponds to the ranges of $\beta$ and $\lambda_{\rm{plug}}$ that can be reached using the full expression of $\mu(I)$ (Eq.~\ref{eqn:muI}). If the simplified expression is used (Eq.~\ref{eqn:muIsimple}), the values of $\lambda_{\rm{plug}}$ are bounded between 0 and 1 and those of 
$\beta$ between $\sqrt{2\mu_s}$ and $+\infty$.
From the two latter equations we can easily show that, in such a condition, the velocity profile tends toward a parabola and the maximum velocity diverges like $\beta^{-1}$.}

\section{Conclusion and discussion}
In this paper, we studied theoretically granular flows in the framework of the $\mu(I)$ rheology. We focused on steady and fully developed granular flows in two geometries: the shear plane and Poiseuille. We obtained  results can be summarized as follow: 

In the shear plane configuration, we have shown that for appropriate parameters, the flow is spatially localized. This is consistent with many experimental observations.
We have also identified a law characterizing the flow, including $ \Lambda_{\rm{critical}} \propto \sqrt {\alpha / \varepsilon} $ [see Eqs. (\ref{eqn:Lambda_epsilon_alpha}) and (\ref{eqn:Lambda_simplify})], i.e. the height $ H $ on which the granular medium is in motion is proportional to $ U_0^{1/2} P_0^{1/4} $.
Although this law is not valid for all values of $ \alpha $ and $ \varepsilon $ the domain of applicability seems very broad. 

In the Poiseuille configuration, we have described in detail the influence of the parameter $ \beta $ which is a function of the pressure gradient in the flow. We have shown that 
{ the granular material flows only if the pressure gradient is greater than a threshold value and that, under certain circumstances, the system behaves like a Bingham plastic fluid.\\
}

\begin{figure}[htbp]
\centering
\includegraphics[width=7cm]{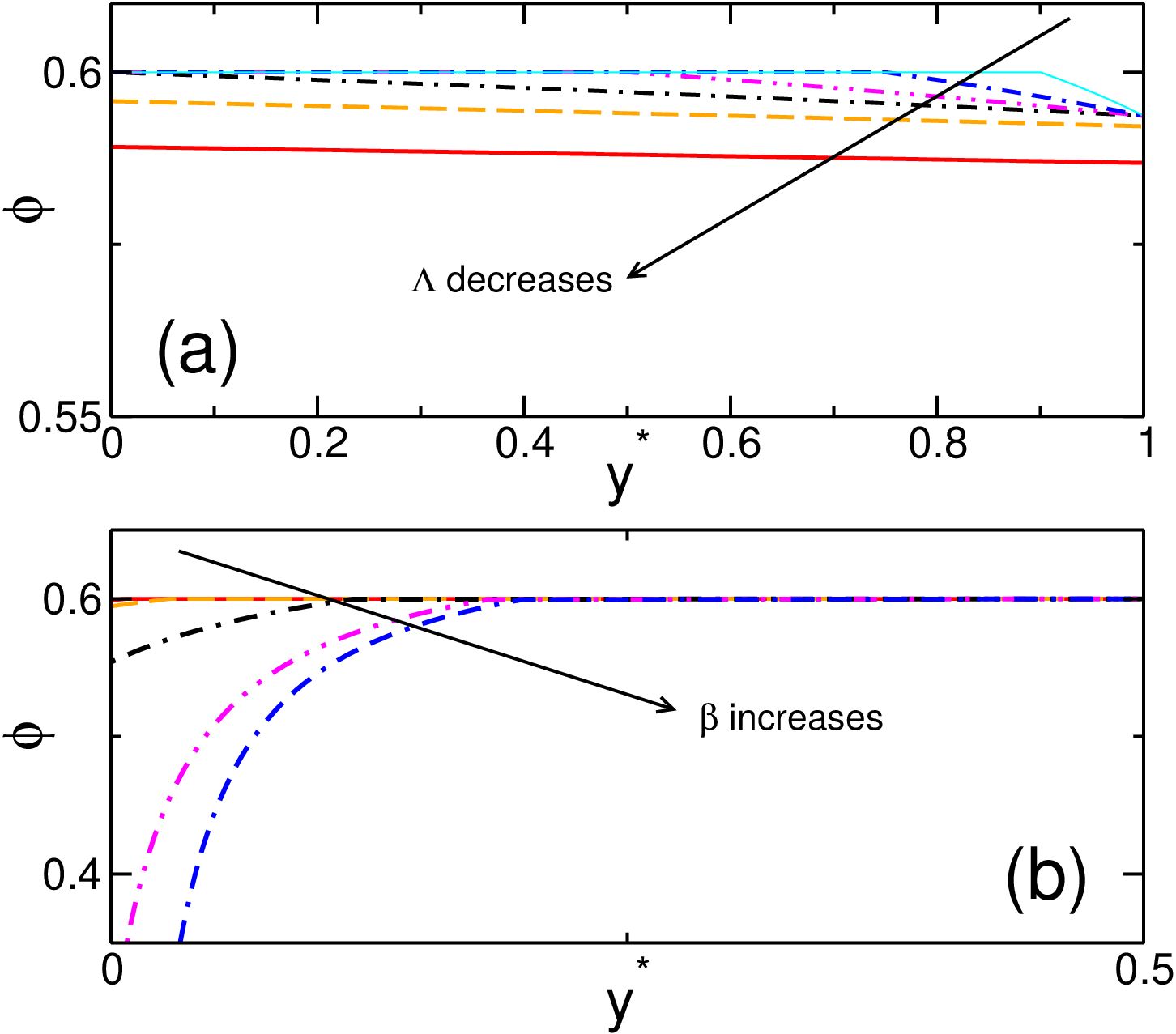}
\caption{The variations of the packing fraction profiles versus $y^*$ is found to be almost constant for the shear plane flow (a). The values of $\Lambda$ are 
$\Lambda=2.5,\ 5,\ 10,\ 20,\ 40 \mbox{ and }100$. On  the contrary, for the Poiseuille flow, it displays important variations close to the sidewalls for important pressure gradients (b).}\label{fig:compa}
\end{figure}

{
As mentioned above, the $\mu(I)$ rheology does not take into account the influence of the fluctuating energy flux that cannot be disregarded close to a boundary, especially when the granular material is not dense. In the case of the plane shear flow, we obtain a qualitative agreement with experiments~\cite{GDR2004} and simulations~\cite{Cruz2004,Richard2008} which suggests that neglecting the energy flux is a reasonable assumption. 
The case of Poiseuille flow is more difficult since few experimental data are available in such a configuration. Our goal was not to compare directly experimental results with the $\mu(I)$ rheology but to apply the latter to a geometry commonly used in fluid mechanics. However the following discussion will shed some light on that particular point.\\
In this work, we have assumed that the packing fraction is constant within the granular material, assumption that can be checked \textit{a posteriori}.
For that purpose, we can use Eq.~\ref{eqn:compa} (with $\zeta=0.2$) and the expressions of $I$ derived in Sect.~\ref{sec:couette} for the shear plane flow and in Sect.~\ref{sec:poiseuille} for the Poiseuille flow 
and see whether  
the variations of $\Phi$ are important or not.
Figure~\ref{fig:compa} reports packing fraction profiles for the shear plane flow (a) and the Poiseuille flow (b). For the former geometry 
the packing fraction varies slightly ($<$ 2\%) justifying the approach used in this work.
This is not surprising since, as mentioned above, our results agree with numerical and experimental results. 
In the Poiseuille geometry, at low $\beta$, the packing fraction variations are also weak. On the contrary, when $\beta$ is increased, low values of the packing fraction are found close to sidewalls. Therefore, the approach used above is no more valid for those conditions ({i.e.} close to boundaries at high values of $\beta$). As mentioned above this was expected since the $\mu(I)$
 rheology does not take into account the energy flux which are important close to the boundaries. 
Note however that the aforementioned conclusions obtained in that geometry at low $\beta$ remain fully valid.}

\section*{Acknowledgments}
We are deeply indebted to J.T.~Jenkins and D.~Berzi for fruitful discussions.  This work
is supported by  the R\'egion Bretagne (CREATE SAMPLEO). M. T. is supported by the R\'egion Bretagne (ARED grant).


\end{document}